\documentclass[a4paper]{article}
\usepackage{INTERSPEECH2020, caption, multirow, makecell, array}
\title{Multi-Task Network for Noise-Robust Keyword Spotting and Speaker Verification using CTC-based Soft VAD and Global Query Attention}
\name{Myunghun Jung, Youngmoon Jung, Jahyun Goo, and Hoirin Kim}
\address{School of Electrical Engineering, KAIST, Daejeon, Republic of Korea}
\email{\{kss2517, dudans, jahyun.goo, hoirkim\}@kaist.ac.kr}
\begin{document}
\maketitle
%
\begin{abstract}
Keyword spotting (KWS) and speaker verification (SV) have been studied independently although it is known that acoustic and speaker domains are complementary.
In this paper, we propose a multi-task network that performs KWS and SV simultaneously to fully utilize the interrelated domain information.
The multi-task network tightly combines sub-networks aiming at performance improvement in challenging conditions such as noisy environments, open-vocabulary KWS, and short-duration SV, by introducing novel techniques of connectionist temporal classification (CTC)-based soft voice activity detection (VAD) and global query attention.
Frame-level acoustic and speaker information is integrated with phonetically originated weights so that forms a word-level global representation.
Then it is used for the aggregation of feature vectors to generate discriminative embeddings.
Our proposed approach shows 4.06\% and 26.71\% relative improvements in equal error rate (EER) compared to the baselines for both tasks.
We also present a visualization example and results of ablation experiments.
\end{abstract}
%
\noindent\textbf{Index Terms}: CTC-based soft VAD, global query attention, keyword spotting, speaker verification, multi-task network 
%
\section{Introduction}
\label{sec:secone}
Progress in speech processing such as speech recognition and text-to-speech enables users to interact with smart devices through a voice user interface (VUI) rather than directly controlling it.
But these techniques mainly focus on spontaneousness during the interaction.
Before that, it is practically important that the interaction starts well.
Among various criteria where the device recognizes the start of the interaction, two typical approaches are keyword spotting and speaker verification.

Keyword spotting (KWS) is the task of detecting the prescribed spoken term in the input utterance.
Many studies and applications have focused on pre-defining the keyword(s) as their product name for wake-up or command words for specific actions \cite{chen2014small, alvarez2019end, tang2018deep}.
Meanwhile, according to users' convenience and customizing needs, some research on open-vocabulary KWS has attracted interest since the users can define any keywords.
A typical way to handle arbitrary keywords is to express any words as acoustic word embeddings which are fixed-dimensional vector representations of arbitrary-length words.
These embeddings learn the acoustic similarity between pronunciations of words pair so that they can encode acoustic information.
In training, some approaches use cross-entropy loss \cite{levin2013fixed, chen2015query}, but triplet loss is mainly used because it can directly map the similarity to the relative distance in embedding space \cite{kamper2016deep, settle2017query, jung2019additional}.
Recently, an approach that considers phonetic information based on connectionist temporal classification (CTC) \cite{graves2006connectionist} together showed good results \cite{lim2019interlayer}.
Still, open-vocabulary KWS has a lot of room for improvement due to its challenging nature.

Speaker verification (SV) is the task of verifying the current speaker is a valid user.
Here, we only deal with text-independent SV that does not have any restrictions on speech contents.
SV requires an enrollment which is a process of registering the user's speaker identity.
Then, speaker information is extracted from each input utterance and compared with the enrolled data.
For successful SV, this speaker information must be expressed as a speaker discriminative representation.
Recent the most powerful approaches based on deep neural networks are encoding speaker information as a fixed-dimensional vector representation, so-called speaker embedding.
For learning discriminative embeddings, the networks are trained to classify speakers using cross-entropy loss \cite{snyder2017deep, snyder2018x} or to group speakers in embedding space using triplet loss \cite{li2017deep, wan2018generalized}.
The criticized problem of these systems is that a long utterance must be used for the input as well as the enrollment to extract speaker information reliably.
It is because the amount of accumulated information increases as the speech lengthens under the assumption that there is one speaker for one utterance.
To cover the problem, several approaches with pooling methods \cite{okabe2018attentive, jung2019spatial, jung2019self} have been proposed to weight the relevant speech frames.
However, if the input length is not long enough, their performances are still degraded.
Accordingly, many short-duration SV studies are being conducted to have high performance even with a short utterance \cite{kanagasundaram2011vector, bhattacharya2017deep, jung2019short, jung2020improving}.

Even though acoustic and speaker information considers each other as a marginal feature that should be suppressed for robust discriminative learning, both KWS and SV have been handled independently.
The ideal situation we think of is that the device can detect the keyword and verify the user at the same time using a short word-level utterance defined by the user.
In other words, open-vocabulary KWS and short-duration SV will eventually operate with the same input in the same conditions.

So in this paper, we propose a multi-task network that performs both KWS and SV simultaneously by fully utilizing acoustic, speaker, and phonetic information.
The multi-task network consists of an enhancement network, acoustic feature extraction network, speaker feature extraction network, and pooling network.
The sub-networks are trained by being shared or contributed to each other.
In this process, we also introduce novel techniques of CTC-based soft voice activity detection (VAD) and global query attention.
We evaluate our proposed approach on discrimination tasks for KWS and SV, respectively.
Experimental results demonstrate that acoustic, speaker, and phonetic domains are interrelated and it is effective to integrate them for learning discriminative embeddings even in noisy environments, open-vocabulary, and short-duration conditions.
Also, we present a visualization example to intuitively understand the proposed methods, and results of ablation experiments to show the effectiveness of the multi-task network.
%
\section{Multi-Task Network}
\label{sec:sectwo}
%
\begin{figure}[t]
\centering
\begin{minipage}[b]{0.55\linewidth}
\includegraphics[width=\linewidth]{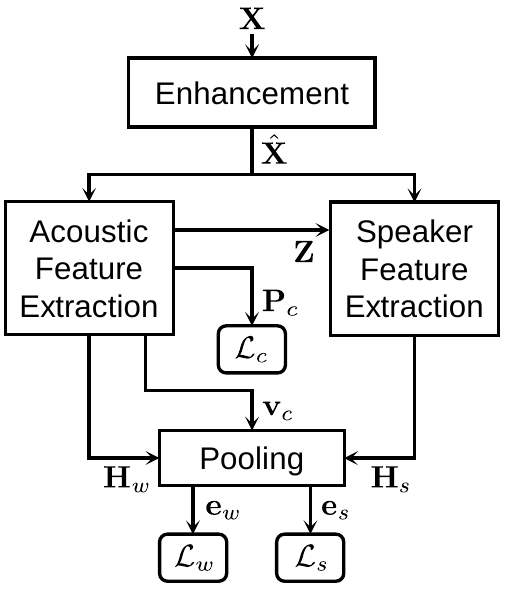}
\end{minipage}
\caption{Overall architecture of the multi-task network.}
\label{fig:fig1}
\end{figure}
\begin{figure}[t]
\begin{minipage}[b]{0.172\linewidth}
\centering
\includegraphics[width=\linewidth]{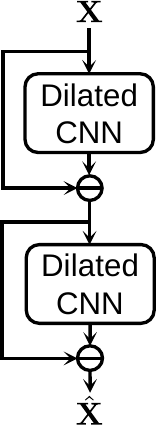}
\caption*{(a)}
\label{fig:fig2a}
\end{minipage}
\hfill
\begin{minipage}[b]{0.317\linewidth}
\centering
\includegraphics[width=\linewidth]{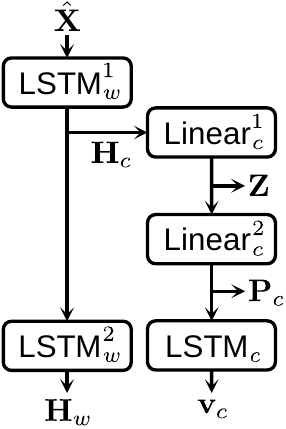}
\caption*{(b)}
\label{fig:fig2b}
\end{minipage}
\hfill
\begin{minipage}[b]{0.25\linewidth}
\centering
\includegraphics[width=\linewidth]{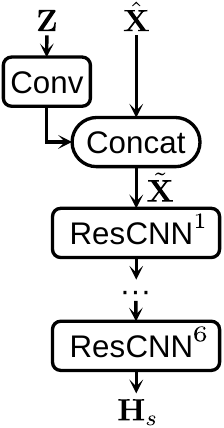}
\caption*{(c)}
\label{fig:fig2c}
\end{minipage}
\centering
\begin{minipage}[b]{0.6\linewidth}
\vspace{3mm}
\includegraphics[width=\linewidth]{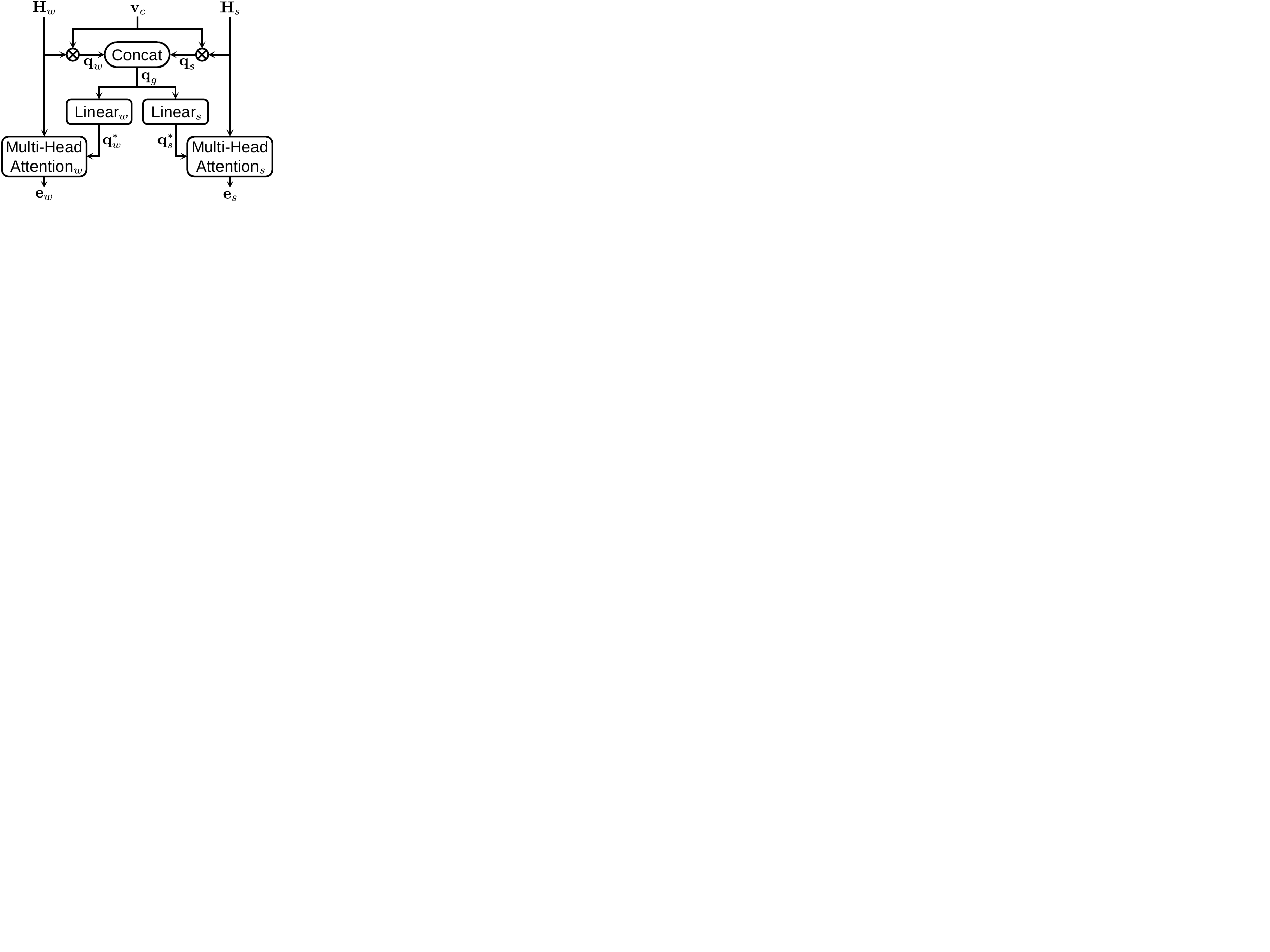}
\caption*{(d)}
\label{fig:fig2d}
\end{minipage}
\caption{Detailed structures of sub-networks: (a) Enhancement network, (b) acoustic feature extraction network, (c) speaker feature extraction network, and (d) pooling network.}
\label{fig:fig2}
\end{figure}

%
In this section, we introduce a multi-task network, which is composed of 4 sub-networks as depicted in Fig.\,\ref{fig:fig1}. The more detailed structures of sub-networks are shown in Fig.\,\ref{fig:fig2}.
%
%
\subsection{Enhancement Network}
\label{ssec:ssectwoone}
With the basic belief that both KWS and SV performance can be improved when the noise component is removed from the input speech, we share the enhancement network with the next two sub-networks.
The enhancement network comprises two dilated convolutional neural network (CNN) cascaded with two residual paths and subtractions (Fig.\,\ref{fig:fig2}.\,(a)).
Each dilated CNN consists of 5 convolution layers and their parameters are noted in Tab.\,\ref{tab:tab1}.
We extract a 256-dimensional log-magnitude spectrogram $\textbf{X}$ with a frame length of 25 ms and a shift of 10 ms yielding a $T \times 256$ input, where $T$ is the number of frames.
The dilated CNN estimates a spectral distortion which is then subtracted from the input.
After two consecutive subtractions, we use the output feature vectors $\hat{\textbf{X}}$ as an enhanced spectrogram.
\subsection{Acoustic Feature Extraction Network}
\label{ssec:ssectwotwo}
In the acoustic feature extraction network, we use two 2-layer bi-directional long short-term memory (LSTM) \cite{hochreiter1997long} modules hierarchically ($\textsf{LSTM}_w^1$ and $\textsf{LSTM}_w^2$ in Fig.\,\ref{fig:fig2}.\,(b)) to represent $\hat{\textbf{X}}$ as frame-level acoustic feature vectors, $\textbf{H}_c \in \mathbb{R}^{T \times 512}$ and $\textbf{H}_w \in \mathbb{R}^{T \times 512}$.
Also, we input $\textbf{H}_c$ into linear layers, $\textsf{Linear}_c^1$ with ReLU activation and $\textsf{Linear}_c^2$ with log-softmax, to capture frame-level phonetic information:
\begin{equation}
\label{eq1}
\textbf{Z} = \textsf{Linear}_c^1 \left( \textbf{H}_c \right ) \in \mathbb{R}^{T \times 256},
\end{equation}
\begin{equation}
\label{eq2}
\textbf{P}_c = \textsf{Linear}_c^2 \left( \textbf{Z} \right) \in \mathbb{R}^{T \times |\mathcal{C}|}.
\end{equation}
$\textbf{P}_c$ is log-probabilities of observing CTC label sequence $\pi = \left( \pi_1, \cdots \pi_T \right)$, where $\pi_t$ is an element of the set $\mathcal{C}$ including characters and blank $\left( | \mathcal{C} | = 27 \right)$.
After trained with CTC loss, $\textbf{P}_c$ becomes a precise indicator of phonetically important frames.
To utilize this property, we tried to transform $\textbf{P}_c$ into the soft VAD \cite{mclaren2015softsad} posteriors.
However, there were problems that the shape of probabilities is too spiky so most of frames are ignored and the blank indicates not only non-speech frames but also the repetition of the previous character.
So we add 1-layer bi-directional LSTM module ($\textsf{LSTM}_c$) to make the distribution more smoother.
The LSTM has $| \mathcal{C} |$ states followed by 1-dimensional projection and sigmoid activation.
We call the output vector $\textbf{v}_c \in \mathbb{R}^T$ as \textit{CTC-based soft VAD} posteriors.
\subsection{Speaker Feature Extraction Network}
\label{ssec:ssectwothree}
We use the bottleneck states $\textbf{Z}$ of Eq.\,\ref{eq1} as phonetic conditional vectors.
To adjust domain mismatch between phonetic and speaker domains, $\textbf{Z}$ enters one convolution layer ($\textsf{Conv}$ in Fig.\,\ref{fig:fig2}.\,(c)), being augmented to 3-channel.
We concatenate it with $\hat{\textbf{X}}$ and get a phonetically conditioned feature vectors $\tilde{\textbf{X}} \in \mathbb{R}^{4 \times T \times 256}$.
The advantage of the phonetic conditioning is that the network can have more knowledge of input speech \cite{zhou2019cnn} even in the short-duration condition so that it becomes easier to suppress unnecessary phonetic variations.
The rest of the network consists of 6 modified version of ResNet \cite{he2016deep}, denoted as $\textsf{ResCNN}^{\ell} \,_{\left( \ell = 1, \cdots , 6 \right)}$.
Each module has one convolution layer and two residual blocks as described in Tab.\,\ref{tab:tab1}.
Since we want to get frame-level speaker information, we does not change the temporal shape.
At the last average and transpose layer, the speaker feature vectors $\textbf{H}_s \in \mathbb{R}^{T \times 256}$ are extracted.
%
\begin{table}[th]
\caption{CNN parameters for the enhancement network and speaker feature extraction network. All convolution layers are followed by batch normalizations and ReLU activations.\\Notation: C- channel, K- kernel size, S- stride, and D- dilation.}
\label{tab:tab1}
\centering
\scriptsize
\begin{tabular}{c|c|c}
\hline
& & \\[-2ex]
Module & Output Size & C, K, S, D \\
\hline
\hline
& & \\[-2ex]
\multirow{3}*{\makecell{\textsf{Dilated}\\ \textsf{CNNs}}} & \multirow{3}*{$1 \times T \times 256$} & $16, (3 \times 3), 1, 1$\\
& & $\big[ 16, (3 \times 3), 1, 2 \big] \times 3$ \\
& & $1, (3 \times 3), 1, 2$ \\
\hline
& & \\[-2ex]
\textsf{Conv} & $3 \times T \times 256$ & $3, (3 \times 3), 1, 1$ \\
\hline
& & \\[-2ex]
\multirow{3}*{$\textsf{ResCNN}^1$} & \multirow{3}*{$8 \times T \times 128$} & $8, (3 \times 3), (1 \times 2), 1$ \\
& & $\Big[ \,\makecell{8, (3 \times 3), 1, 1 \\ 8, (3 \times 3), 1, 1} \,\Big] \times 2$ \\
\hline
& & \\[-2ex]
\multirow{3}*{$\textsf{ResCNN}^2$} & \multirow{3}*{$16 \times T \times 64$} & $16, (3 \times 3), (1 \times 2), 1$ \\
& & $\Big[ \,\makecell{16, (3 \times 3), 1, 1 \\ 16, (3 \times 3), 1, 1} \,\Big] \times 2$ \\
\hline
& & \\[-2ex]
\multirow{3}*{$\textsf{ResCNN}^3$} & \multirow{3}*{$32 \times T \times 32$} & $32, (3 \times 3), (1 \times 2), 1$ \\
& & $\Big[ \,\makecell{32, (3 \times 3), 1, 1 \\ 32, (3 \times 3), 1, 1} \,\Big] \times 2$ \\
\hline
& & \\[-2ex]
\multirow{3}*{$\textsf{ResCNN}^4$} & \multirow{3}*{$64 \times T \times 16$} & $64, (3 \times 3), (1 \times 2), 1$ \\
& & $\Big[ \,\makecell{64, (3 \times 3), 1, 1 \\ 64, (3 \times 3), 1, 1} \,\Big] \times 2$ \\
\hline
& & \\[-2ex]
\multirow{3}*{$\textsf{ResCNN}^5$} & \multirow{3}*{$128 \times T \times 8$} & $128, (3 \times 3), (1 \times 2), 1$ \\
& & $\Big[ \,\makecell{128, (3 \times 3), 1, 1 \\ 128, (3 \times 3), 1, 1} \,\Big] \times 2$ \\
\hline
& & \\[-2ex]
\multirow{4}*{$\textsf{ResCNN}^6$} & \multirow{3}*{$256 \times T \times 4$} & $256, (3 \times 3), (1 \times 2), 1$ \\
& & $\Big[ \,\makecell{256, (3 \times 3), 1, 1 \\ 256, (3 \times 3), 1, 1} \,\Big] \times 2$ \\
\cline{2-3}
& & \\[-2ex]
& $T \times 256$ & average, transpose \\
\hline
\end{tabular}
\end{table}
%
\begin{figure}[t]
\begin{minipage}[b]{0.48\linewidth}
\centering
\includegraphics[width=\linewidth]{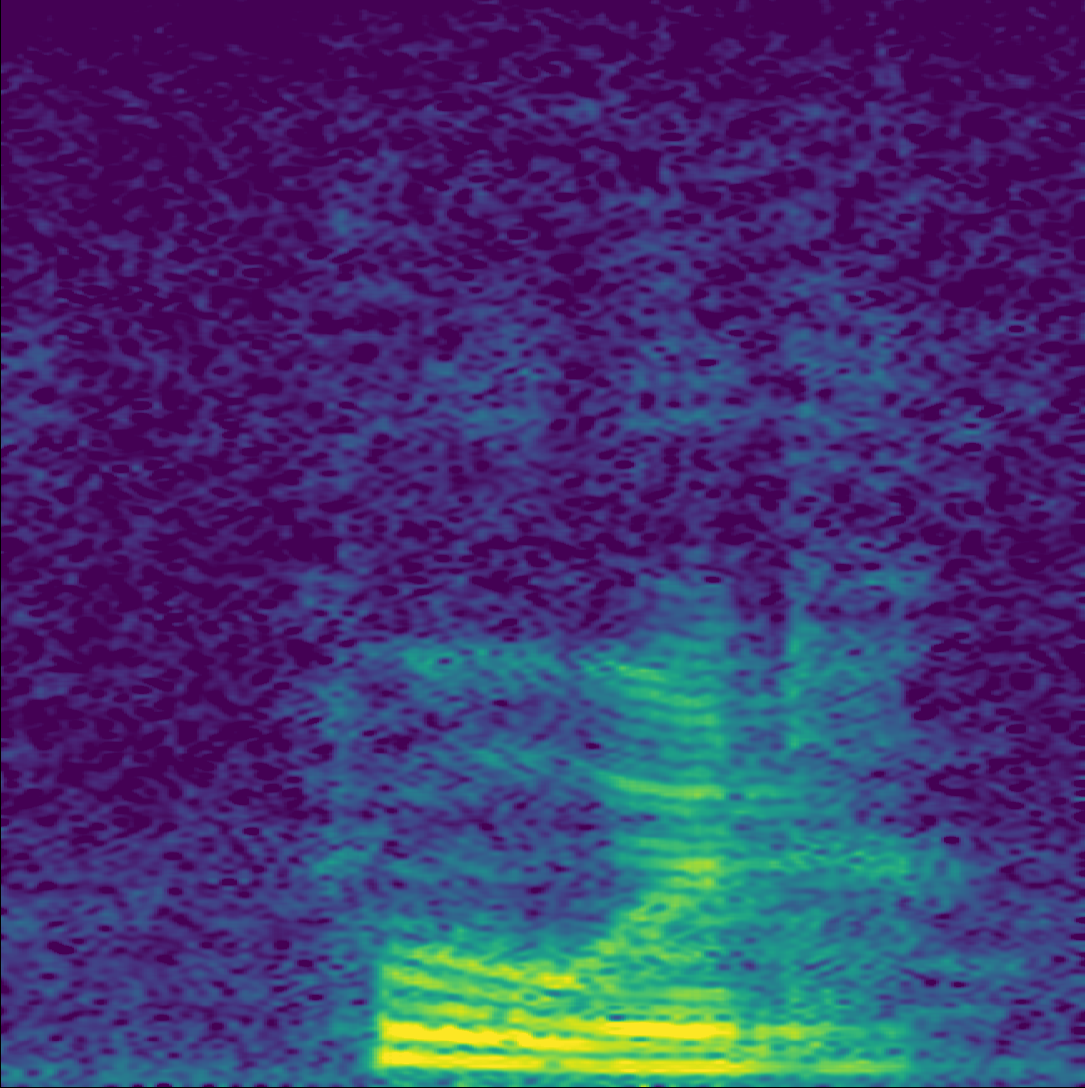}
\caption*{(a) $\textbf{X}_{clean}$}
\label{fig:fig3a}
\end{minipage}
\hfill
\begin{minipage}[b]{0.48\linewidth}
\centering
\includegraphics[width=\linewidth]{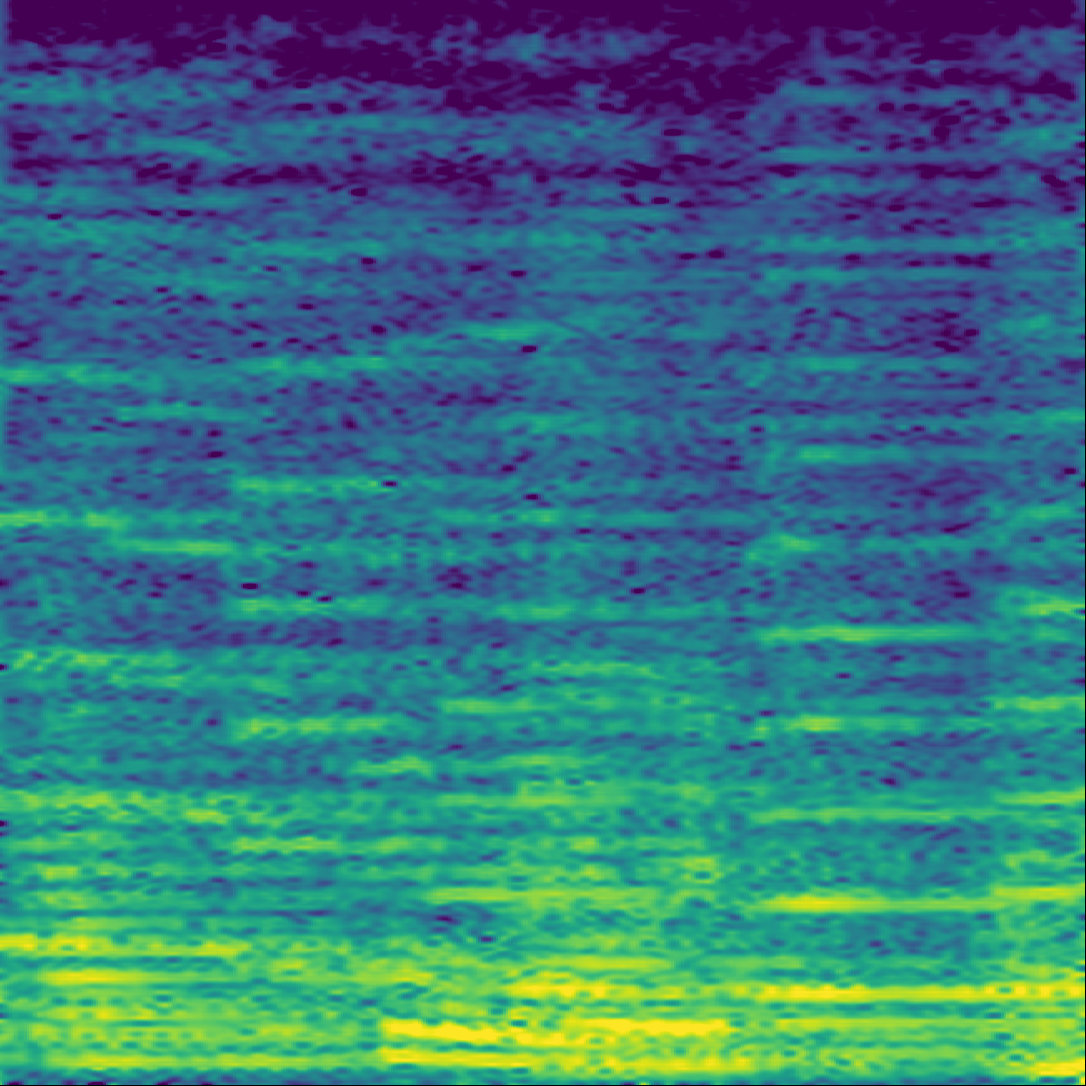}
\caption*{(b) $\textbf{X}_{noisy}$}
\label{fig:fig3b}
\end{minipage}
\begin{minipage}[b]{0.48\linewidth}
\centering
\vspace{3mm}
\includegraphics[width=\linewidth]{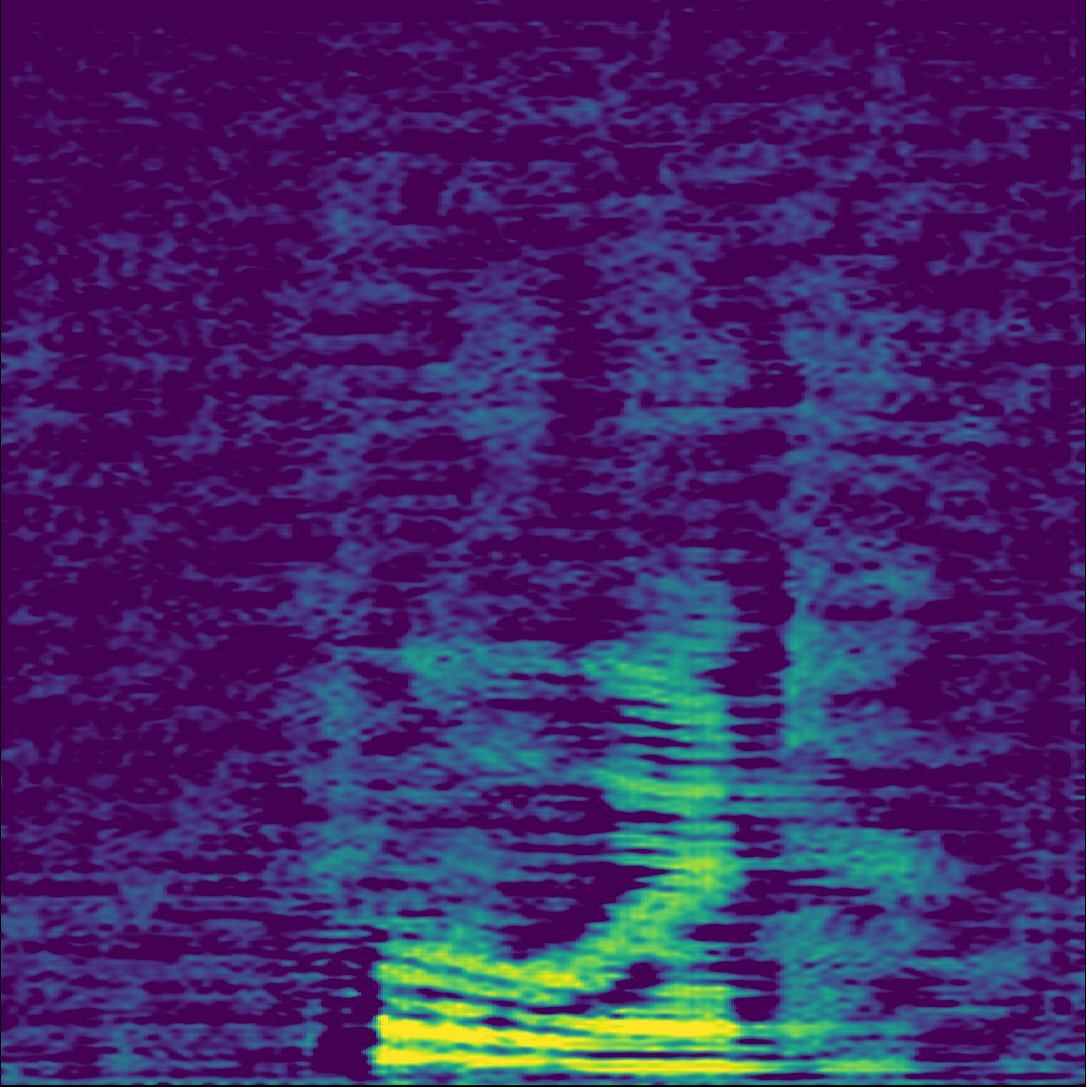}
\caption*{(c) $\hat{\textbf{X}}_{clean}$}
\label{fig:fig3c}
\end{minipage}
\hfill
\begin{minipage}[b]{0.48\linewidth}
\centering
\includegraphics[width=\linewidth]{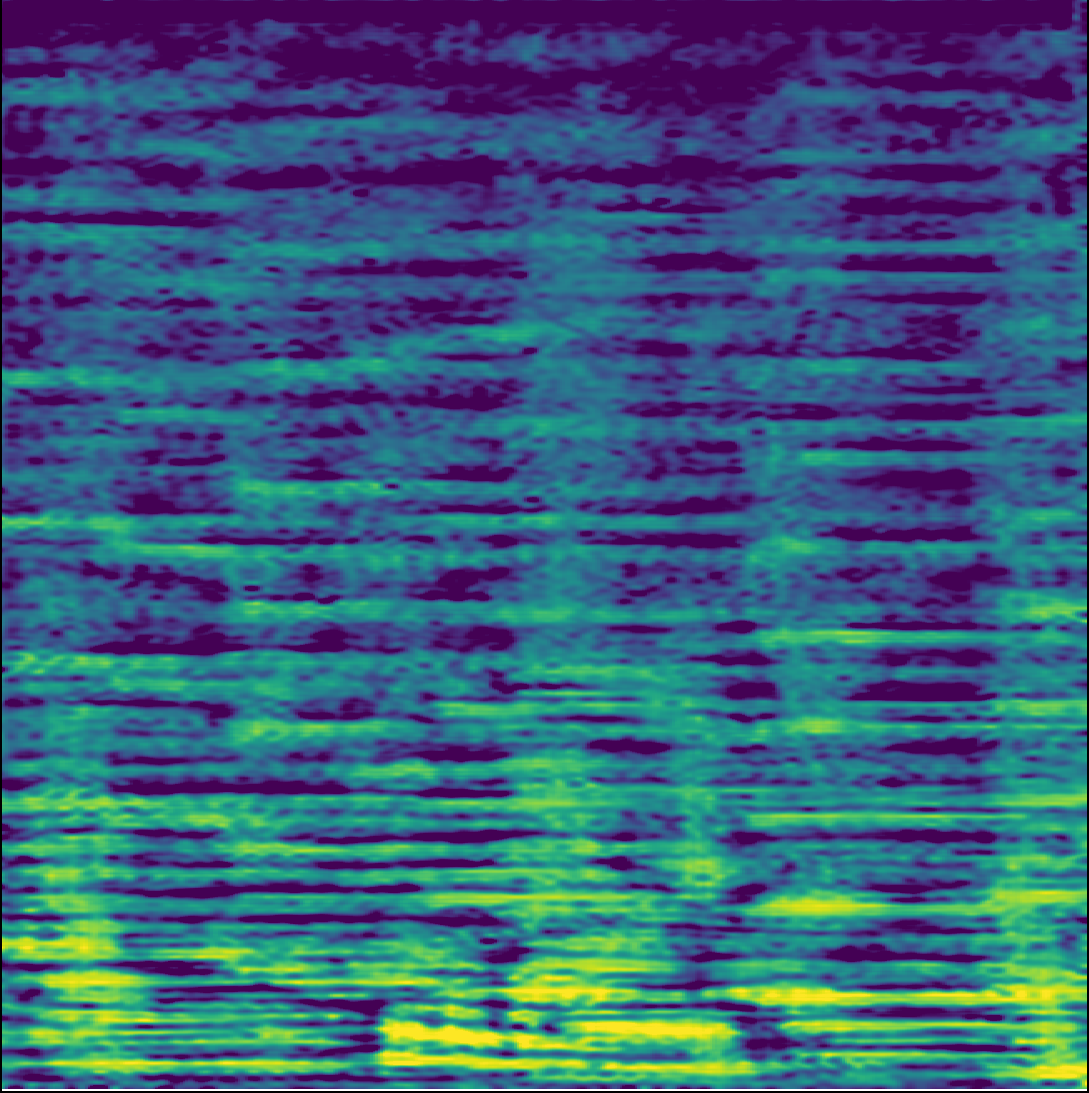}
\caption*{(d) $\hat{\textbf{X}}_{noisy}$}
\label{fig:fig3d}
\end{minipage}
\begin{minipage}[b]{0.48\linewidth}
\centering
\vspace{3mm}
\frame{\includegraphics[width=\linewidth]{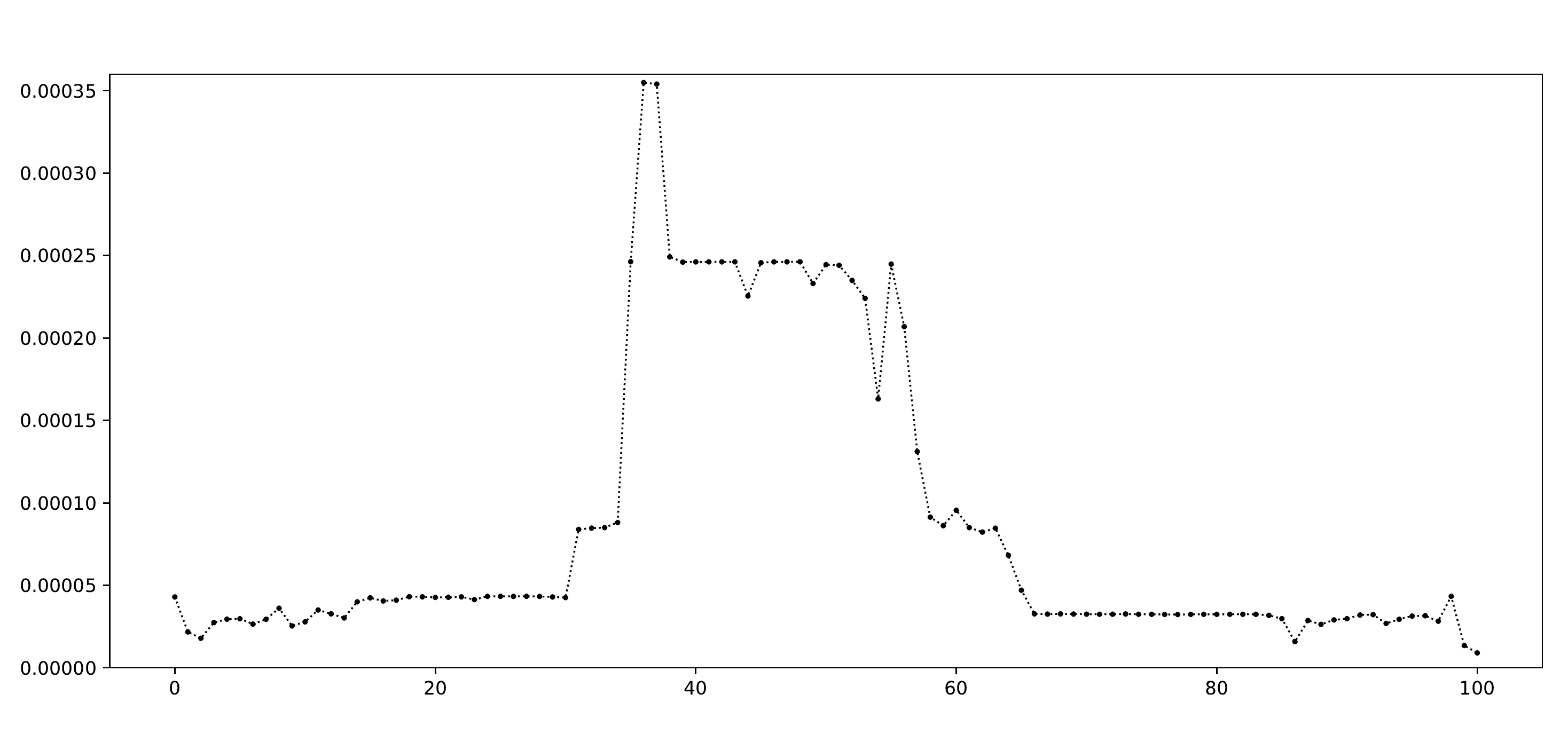}}
\caption*{(e) $\textbf{v}_{c, clean}$}
\label{fig:fig3e}
\end{minipage}
\hfill
\begin{minipage}[b]{0.48\linewidth}
\centering
\frame{\includegraphics[width=\linewidth]{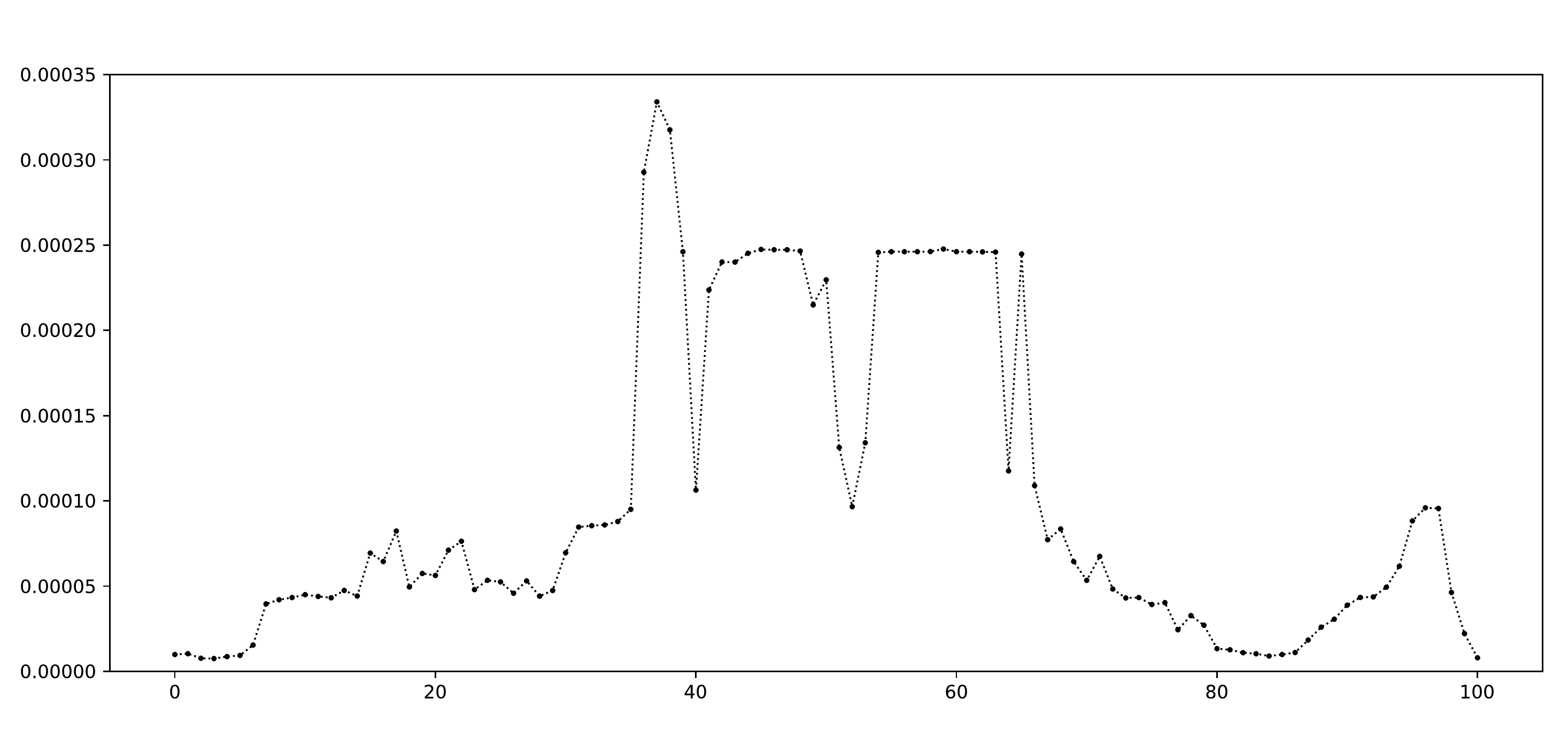}}
\caption*{(f) $\textbf{v}_{c, noisy}$}
\label{fig:fig3f}
\end{minipage}

\caption{Visualization of one utterance in the test set. The first column is for clean speech and the second is for its corrupted version with Music noise at the SNR of 0 dB. (a)-(b): Original spectrograms. (c)-(d): Enhanced spectrograms. (e)-(f): CTC-based soft VAD posteriors. (Recommend viewing in color)}
\label{fig:fig3}
\end{figure}
%
\subsection{Pooling Network}
\label{ssec:ssectwofour}
Here, we propose a novel pooling method with \textit{global query attention}, that integrates all domain information at the query generation stage.
The method is composed of 4 steps, computing temporary queries with CTC-based soft VAD, global query generation, recomputing mutually informed domain queries, and aggregation.
The temporary queries $\textbf{q}_w$ and $\textbf{q}_s$ in Eq.\,\ref{eq3} are average representatives corresponding to phonetically relevant speech frames.
Specifically, these are results of weighted sum of feature vectors using $\textbf{v}_c$ as the frame-wise weights.
Then we generate a global query $\textbf{q}_g$ by concatenating two temporary queries as:
\begin{equation}
\label{eq3}
\textbf{q}_{g} = \left[ \textbf{q}_w , \textbf{q}_s \right] = \left[ \textbf{H}_{w}^T \textbf{v}_c , \textbf{H}_s^T \textbf{v}_c \right]  \in \mathbb{R}^{d_w + d_s} .
\end{equation}
As a result, the global query contains not only both acoustic and speaker information of the input utterance but also phonetic information in that temporary queries are derived with phonetically originated weights $\textbf{v}_c$.

Since the ratio at which the global query contributes to each domain is different, the global query goes through two linear layers in parallel ($\textsf{Linear}_w$ and $\textsf{Linear}_s$ in Fig.\,\ref{fig:fig2}.\,(d)) to recompute mutually informed domain queries $\textbf{q}_w^\ast$ and $\textbf{q}_s^\ast$.

We adopt the multi-head attention scheme \cite{vaswani2017attention} for the aggregation, where the key and value are same as the acoustic or speaker feature vectors and they are aggregated with the mutually informed domain query:
\begin{equation}
\label{eq4}
\textbf{q}_k^\ast = \textsf{Linear}_k \left( \textbf{q}_g \right) \in \mathbb{R}^{d_k} ,
\end{equation}
\begin{equation}
\label{eq5}
\textbf{e}_k = \textsf{Multi-Head}_k \left( \textbf{q}_k^\ast , \textbf{H}_k , \textbf{H}_k \right) \in \mathbb{R}^{d_k} ,
\end{equation}
where $k \in \{ w, s \}$, $d_w = 512$, $d_s = 256$, and we employ 4 heads.
The resulting two vectors $\textbf{e}_w$ and $\textbf{e}_s$ are the acoustic word embedding and speaker embedding, respectively.
\subsection{Loss Function}
\label{ssec:ssectwofive}
As depicted in Fig.\,\ref{fig:fig1}, we use 3 simple loss functions.
Two $L_2$ constrained softmax losses \cite{ranjan2017l2} $\mathcal{L}_w$ and $\mathcal{L}_s$ are used for KWS and SV, where $\left \| \textbf{e}_w \right \| = 6$ and $\left \| \textbf{e}_s \right \| = 12$, and $\mathcal{L}_c$ is the CTC loss.
The overall training loss is $\mathcal{L} = \mathcal{L}_w + \mathcal{L}_s + \mathcal{L}_c$.
We do not use any explicit loss function for the enhancement purpose.
%
\begin{table*}[th]
\caption{EER (\%) results on the test set. (-) and (+) denote the exclusion and substitution for the ablation, respectively.}
\label{tab:tab2}
\centering
\scriptsize
\begin{tabular}{c|c||>{\raggedleft}p{0.7cm}@{\hspace{2pt}}|>{\raggedleft}p{0.7cm}@{\hspace{2pt}}||>{\raggedleft}p{0.7cm}@{\hspace{2pt}}|>{\raggedleft}p{0.7cm}@{\hspace{2pt}}||>{\raggedleft}p{0.8cm}@{\hspace{2pt}}|>{\raggedleft}p{0.8cm}@{\hspace{2pt}}||>{\raggedleft}p{0.8cm}@{\hspace{2pt}}|>{\raggedleft}p{0.8cm}@{\hspace{2pt}}||>{\raggedleft}p{0.9cm}@{\hspace{2pt}}|>{\raggedleft}p{0.9cm}@{\hspace{2pt}}||>{\raggedleft}p{0.7cm}@{\hspace{2pt}}|>{\raggedleft}p{0.7cm}@{\hspace{1pt}}c@{\hspace{1pt}}}
\hline
\multicolumn{2}{c||}{} & \multicolumn{2}{c||}{Proposed} & \multicolumn{2}{c||}{Baselines} & \multicolumn{2}{c||}{\makecell{(-) Enhancement\\network}} & \multicolumn{2}{c||}{\makecell{(-) Phonetic\\conditioning}} & \multicolumn{2}{c||}{\makecell{(-) Pooling network\\(+) Self-attention}} & \multicolumn{2}{c}{\makecell{(-) CTC}} & \\
\hline
& & & & & & & & & & & & & & \\[-2ex]
Type & SNR & \multicolumn{1}{c|}{KWS} & \multicolumn{1}{c||}{SV} & \multicolumn{1}{c|}{KWS} & \multicolumn{1}{c||}{SV} & \multicolumn{1}{c|}{KWS} & \multicolumn{1}{c||}{SV} & \multicolumn{1}{c|}{KWS} & \multicolumn{1}{c||}{SV} & \multicolumn{1}{c|}{KWS} & \multicolumn{1}{c||}{SV} & \multicolumn{1}{c|}{KWS} & \multicolumn{1}{c}{SV} & \\
\hline
\hline
\multicolumn{2}{c||}{} & & & & & & & & & & & & & \\[-2ex]
\multicolumn{2}{c||}{Clean} & \textbf{4.52} & 2.47 & 4.70 & 2.76 & 4.89 & 2.44 & 4.77 & \textbf{2.33} & 5.01 & 3.83 & 5.72 & 5.57 & \\
\hline
\hline
& & & & & & & & & & & & & & \\[-2ex]
\multirow{5}{*}{Music} & 20 & \textbf{4.72} & \textbf{4.24} & 4.91 & 6.15 & 5.45 & 4.63 & 5.00 & 4.51 & 5.22 & 6.05 & 5.94 & 6.45 & \\
& 10 & \textbf{5.33} & \textbf{5.59} & 5.58 & 8.42 & 6.51 & 6.18 & 5.67 & 5.89 & 5.89 & 7.59 & 6.82 & 7.95 & \\
& 5 & \textbf{6.40} & \textbf{6.97} & 6.64 & 10.25 & 7.99 & 7.81 & 6.79 & 7.28 & 7.04 & 9.28 & 8.29 & 9.83 & \\
& 0 & \textbf{8.61} & \textbf{9.71} & 8.90 & 13.65 & 11.82 & 10.95 & 9.05 & 9.99 & 9.67 & 12.96 & 11.33 & 13.47 & \\
\cline{2-15}
& & & & & & & & & & & & & & \\[-2ex]
& Avg. & \textbf{6.26} & \textbf{6.63} & 6.51 & 9.62 & 7.94 & 7.39 & 6.63 & 6.92 & 6.96 & 8.97 & 8.09 & 9.43 & \\
\hline
\hline
& & & & & & & & & & & & & & \\[-2ex]
\multirow{5}{*}{Babble} & 20 & \textbf{5.12} & 3.56 & 5.19 & 4.60 & 5.67 & \textbf{3.51} & 5.43 & 3.56 & 5.46 & 5.03 & 6.54 & 6.13 & \\
& 10 & \textbf{6.67} & \textbf{4.61} & 6.92 & 6.32 & 7.98 & 4.63 & 7.16 & 4.63 & 7.55 & 6.33 & 8.70 & 7.66 & \\
& 5 & \textbf{9.11} & \textbf{5.90} & 9.73 & 8.11 & 11.46 & 5.96 & 9.82 & 5.91 & 10.43 & 8.10 & 11.39 & 9.56 & \\
& 0 & \textbf{14.57} & 8.81 & 15.42 & 12.15 & 18.47 & 8.81 & 15.92 & \textbf{8.63} & 16.51 & 12.41 & 16.87 & 13.64 & \\
\cline{2-15}
& & & & & & & & & & & & & & \\[-2ex]
& Avg. & \textbf{8.87} & 5.72 & 9.31 & 7.80 & 10.89 & 5.73 & 9.58 & \textbf{5.68} & 9.98 & 7.97 & 10.87 & 9.25 & \\
\hline
\hline
& & & & & & & & & & & & & & \\[-2ex]
\multirow{5}{*}{Others} & 20 & \textbf{4.70} & \textbf{6.82} & 5.04 & 10.40 & 5.65 & 7.55 & 4.99 & 7.29 & 5.20 & 8.08 & 6.04 & 8.22 & \\
& 10 & \textbf{5.39} & \textbf{9.64} & 5.70 & 13.58 & 6.73 & 10.81 & 5.64 & 10.54 & 5.84 & 10.46 & 6.69 & 10.86 & \\
& 5 & \textbf{6.24} & \textbf{11.99} & 6.36 & 15.87 & 8.27 & 13.56 & 6.29 & 13.34 & 6.95 & 12.91 & 7.69 & 13.41 & \\
& 0 & \textbf{8.00} & \textbf{15.81} & 8.26 & 19.44 & 11.66 & 18.18 & 8.09 & 18.16 & 9.03 & 17.45 & 9.84 & 17.73 & \\
\cline{2-15}
& & & & & & & & & & & & & & \\[-2ex]
& Avg. & \textbf{6.08} & \textbf{11.06} & 6.34 & 14.82 & 8.08 & 12.52 & 6.25 & 12.33 & 6.75 & 12.22 & 7.56 & 12.55 & \\
\hline
\hline
\multicolumn{2}{c||}{} & & & & & & & & & & & & & \\[-2ex]
\multicolumn{2}{c||}{Total Avg.} & \textbf{7.07} & \textbf{7.80} & 7.39 & 10.74 & 8.97 & 8.55 & 7.49 & 8.31 & 7.90 & 9.72 & 8.84 & 10.41 & \\
\hline
\end{tabular}
\end{table*}
%
\section{Experiments}
\label{sec:secthree}
\subsection{Datasets}
\label{ssec:ssecthreeone}
We use Google's Speech Commands dataset V2 \cite{warden2018speech} which consists of 105829 utterances of 35 words spoken by 2618 speakers.
All utterances have a duration of equal to or less than 1 s.
We use only 2118 speakers, excluding those with less than 10 utterances.
Then disjoint sets of 1959 and 159 speakers are randomly selected for training set and test set.
Also, to check the robustness of KWS against unseen words, utterances corresponding to the three words `happy', `marvin', and `sheila' are excluded from the training set.

Since the original dataset was collected using crowdsourcing, the dataset is not perfectly clean, but we regard it as the clean.
For more challenging experiments, we corrupt the original dataset with the three types of noise, `Music', `Babble', and `Others' from the MUSAN dataset \cite{snyder2015musan}.
For the training set, each utterance is augmented with a randomly selected type of noise under the SNR randomly chosen from the set \{20, 10, 5, 0\} (in dB).
For the test set, all types of noise and SNRs are considered, resulting in 13 environments.
\subsection{Experiment Setting}
\label{ssec:ssecthreetwo}
All the networks are implemented with PyTorch \cite{paszke2019pytorch}.
We use the stochastic gradient descent with learning rate of 0.005 and momentum of 0.9 for the enhancement and speaker feature extraction networks.
And the acoustic feature extraction and pooling networks are optimized by the Adam \cite{kingma2015adam} with learning rate of 0.0005.
The mini-batch size is set to 128 and two GTX 1080 Ti GPUs are used.
The multi-task network is trained for 100 epochs and the model of the last epoch is used for evaluation.
\subsection{Evaluation Tasks and Metrics}
\label{ssec:ssecthreethree}
Keyword spotting and speaker verification are basically discrimination tasks that make a decision given a score between embeddings of enrollment and test utterances.
For both tasks, we use the cosine distance to measure the score and performances are evaluated using the equal error rate (EER).
Here all utterances in the test set are used once for enrollment, one by one.
Accordingly, 1 s \,-\, 1 s short-duration constraint is applied.
\subsection{Baselines}
\label{ssec:ssecthreefour}
For performance comparison, we use individually trained networks as the baselines of KWS and SV.
For KWS baseline, the enhancement network and acoustic feature extraction network are used.
This LSTM-based architecture has been used in \cite{chen2015query, settle2017query, jung2019additional}.
Likewise, for SV baseline, the enhancement network and speaker feature extraction network are used without concatenating phonetic condition vectors $\textbf{Z}$.
The remaining network architecture is similar to the basic ResNet-based SV network \cite{he2016deep}.
To aggregate feature vectors, instead of the global query attention, we use the self-attention \cite{lin2017structured} method in Eq.\,\ref{eq6} which has been widely used.
\begin{equation}
\label{eq6}
\textbf{e}_k = \frac{1}{4}\sum_{h=1}^4 \textsf{Softmax} \left( \textbf{V}_k^h \textsf{tanh} \left( \textbf{W}_k \textbf{H}_k^T \right) \right) \textbf{H}_k ,
\end{equation}
where $k \in \{ w, s \}$, $\textbf{W}_k \in \mathbb{R}^{128 \times d_k}$, and $\textbf{V}_k^h \in \mathbb{R}^{128}$.
\subsection{Results}
\label{ssec:ssecthreefive}
In Tab.\,\ref{tab:tab2}, we compare the performance on the test set between our proposed approach and baselines.
Definitely, the proposed approach outperforms the baselines in all environments.
For KWS, there is a slight increment of relatively 4.06\% on average. 
However, for SV, we can obtain high improvements of relatively 26.71\% on average and absolutely 2-4\%.
From this, we demonstrate that acoustic, phonetic, and speaker information can be fully utilized by joint learning the multi-task network.

In Fig.\,\ref{fig:fig3}, we visualize example spectrograms and CTC-based soft VAD posteriors.
We can observe that the enhancement network highlights the harmonics by reducing frequency components on valleys rather than eliminating noise itself, which can increase the frequency resolution of the spectrogram.
Also, noticeable noise reduction at the speech boundaries can be confirmed, which is clearly seen in Fig.\,\ref{fig:fig3}.\,(c).
Likewise, in Fig.\,\ref{fig:fig3}.\,(d), there are narrow but deep divisions, particularly on the lower frequency region.
When looking at $\textbf{v}_c$, there are high posteriors on actual speech frames and it means that very meaningful temporary queries and global query can be generated at the global query attention.

In additional columns in Tab.\,\ref{tab:tab2}, we investigate the effectiveness of the multi-task network with some ablation experiments.
When the enhancement network is excluded, the results of KWS are particularly degraded on low SNRs.
The exception of the phonetic conditioning shows the difference for SV performance on `Others' noise type.
It means that the phonetic information helps frame-level speaker feature extraction even with non-contextual noise.
Meanwhile, the SV result in 0 dB `Babble' indicates that mistaken phonetic information from background speech could make ambiguity.
The next two columns show the global query attention performs better than the self-attention.
Lastly, all the CTC related modules are removed, that is only the enhancement network is shared between two feature extraction networks.
The results imply that the mutual contribution at the feature extractions and pooling network is equally or more effective than the existence of the enhancement network.
%
\section{Conclusion}
\label{sec:secfour}
In this paper, we propose a multi-task network comprising multiple sub-networks.
Also we introduce novel techniques of CTC-based soft VAD and global query attention to tightly utilize interrelated domain information even in challenging conditions of noisy environments, open-vocabulary KWS, and short-duration SV.
Each sub-network is originally designed for individual purposes of enhancement, KWS, and SV, but great performance improvement can be achieved by effectively shared and mutually contributed.
Experimental results demonstrate that the proposed approach outperforms the baselines.
%
\section{Acknowledgement}
\label{sec:secfive}
This work was conducted by Center for Applied Research in Artificial Intelligence(CARAI) grant funded by DAPA and ADD (UD190031RD).
%
%
\bibliographystyle{IEEEtran}
\bibliography{mybib}

\begin{thebibliography}{10}
\providecommand{\url}[1]{#1}
\csname url@samestyle\endcsname
\providecommand{\newblock}{\relax}
\providecommand{\bibinfo}[2]{#2}
\providecommand{\BIBentrySTDinterwordspacing}{\spaceskip=0pt\relax}
\providecommand{\BIBentryALTinterwordstretchfactor}{4}
\providecommand{\BIBentryALTinterwordspacing}{\spaceskip=\fontdimen2\font plus
\BIBentryALTinterwordstretchfactor\fontdimen3\font minus
  \fontdimen4\font\relax}
\providecommand{\BIBforeignlanguage}[2]{{%
\expandafter\ifx\csname l@#1\endcsname\relax
\typeout{** WARNING: IEEEtran.bst: No hyphenation pattern has been}%
\typeout{** loaded for the language `#1'. Using the pattern for}%
\typeout{** the default language instead.}%
\else
\language=\csname l@#1\endcsname
\fi
#2}}
\providecommand{\BIBdecl}{\relax}
\BIBdecl

\bibitem{chen2014small}
G.~Chen, C.~Parada, and G.~Heigold, ``Small-footprint keyword spotting using
  deep neural networks,'' in \emph{Proc. of IEEE International Conference on
  Acoustics, Speech and Signal Processing (ICASSP)}, 2014, pp. 4087--4091.

\bibitem{alvarez2019end}
R.~Alvarez and H.-J. Park, ``End-to-end streaming keyword spotting,'' in
  \emph{Proc. of IEEE International Conference on Acoustics, Speech and Signal
  Processing (ICASSP)}, 2019, pp. 6336--6340.

\bibitem{tang2018deep}
R.~Tang and J.~Lin, ``Deep residual learning for small-footprint keyword
  spotting,'' in \emph{Proc. of IEEE International Conference on Acoustics,
  Speech and Signal Processing (ICASSP)}, 2018, pp. 5484--5488.

\bibitem{levin2013fixed}
K.~Levin, K.~Henry, A.~Jansen, and K.~Livescu, ``Fixed-dimensional acoustic
  embeddings of variable-length segments in low-resource settings,'' in
  \emph{Proc. of IEEE Automatic Speech Recognition and Understanding Workshop
  (ASRU)}, 2013, pp. 410--415.

\bibitem{chen2015query}
G.~Chen, C.~Parada, and T.~N. Sainath, ``Query-by-example keyword spotting
  using long short-term memory networks,'' in \emph{Proc. of IEEE International
  Conference on Acoustics, Speech and Signal Processing (ICASSP)}, 2015, pp.
  5236--5240.

\bibitem{kamper2016deep}
H.~Kamper, W.~Wang, and K.~Livescu, ``Deep convolutional acoustic word
  embeddings using word-pair side information,'' in \emph{Proc. of IEEE
  International Conference on Acoustics, Speech and Signal Processing
  (ICASSP)}, 2016, pp. 4950--4954.

\bibitem{settle2017query}
S.~Settle, K.~Levin, H.~Kamper, and K.~Livescu, ``Query-by-example search with
  discriminative neural acoustic word embeddings,'' in \emph{Proc. of Annual
  Conference of the International Speech Communication Association
  (INTERSPEECH)}, 2017, pp. 2874--2878.

\bibitem{jung2019additional}
M.~Jung, H.~Lim, J.~Goo, Y.~Jung, and H.~Kim, ``Additional shared decoder on
  siamese multi-view encoders for learning acoustic word embeddings,'' in
  \emph{Proc. of IEEE Automatic Speech Recognition and Understanding Workshop
  (ASRU)}, 2019, pp. 629--636.

\bibitem{graves2006connectionist}
A.~Graves, S.~Fern{\'a}ndez, F.~Gomez, and J.~Schmidhuber, ``Connectionist
  temporal classification: labelling unsegmented sequence data with recurrent
  neural networks,'' in \emph{Proc. of International Conference on Machine
  Learning (ICML)}, 2006, pp. 369--376.

\bibitem{lim2019interlayer}
H.~Lim, Y.~Kim, J.~Goo, and H.~Kim, ``Interlayer selective attention network
  for robust personalized wake-up word detection,'' \emph{IEEE Signal
  Processing Letters}, vol.~27, pp. 126--130, 2020.

\bibitem{snyder2017deep}
D.~Snyder, D.~Garcia-Romero, D.~Povey, and S.~Khudanpur, ``Deep neural network
  embeddings for text-independent speaker verification,'' in \emph{Proc. of
  Annual Conference of the International Speech Communication Association
  (INTERSPEECH)}, 2017, pp. 999--1003.

\bibitem{snyder2018x}
D.~Snyder, D.~Garcia-Romero, G.~Sell, D.~Povey, and S.~Khudanpur, ``X-vectors:
  Robust dnn embeddings for speaker recognition,'' in \emph{Proc. of IEEE
  International Conference on Acoustics, Speech and Signal Processing
  (ICASSP)}, 2018, pp. 5329--5333.

\bibitem{li2017deep}
C.~Li, X.~Ma, B.~Jiang, X.~Li, X.~Zhang, X.~Liu, Y.~Cao, A.~Kannan, and Z.~Zhu,
  ``Deep speaker: an end-to-end neural speaker embedding system,'' \emph{arXiv
  preprint arXiv:1705.02304}, 2017.

\bibitem{wan2018generalized}
L.~Wan, Q.~Wang, A.~Papir, and I.~L. Moreno, ``Generalized end-to-end loss for
  speaker verification,'' in \emph{Proc. of IEEE International Conference on
  Acoustics, Speech and Signal Processing (ICASSP)}, 2018, pp. 4879--4883.

\bibitem{okabe2018attentive}
K.~Okabe, T.~Koshinaka, and K.~Shinoda, ``Attentive statistics pooling for deep
  speaker embedding,'' in \emph{Proc. of Annual Conference of the International
  Speech Communication Association (INTERSPEECH)}, 2018, pp. 2252--2256.

\bibitem{jung2019spatial}
Y.~Jung, Y.~Kim, H.~Lim, Y.~Choi, and H.~Kim, ``Spatial pyramid encoding with
  convex length normalization for text-independent speaker verification,'' in
  \emph{Proc. of Annual Conference of the International Speech Communication
  Association (INTERSPEECH)}, 2019, pp. 4030--4034.

\bibitem{jung2019self}
Y.~Jung, Y.~Choi, and H.~Kim, ``Self-adaptive soft voice activity detection
  using deep neural networks for robust speaker verification,'' in \emph{Proc.
  of IEEE Automatic Speech Recognition and Understanding Workshop (ASRU)},
  2019, pp. 365--372.

\bibitem{kanagasundaram2011vector}
A.~Kanagasundaram, R.~Vogt, D.~B. Dean, S.~Sridharan, and M.~W. Mason,
  ``I-vector based speaker recognition on short utterances,'' in \emph{Proc. of
  Annual Conference of the International Speech Communication Association
  (INTERSPEECH)}, 2011, pp. 2341--2344.

\bibitem{bhattacharya2017deep}
G.~Bhattacharya, M.~J. Alam, and P.~Kenny, ``Deep speaker embeddings for
  short-duration speaker verification,'' in \emph{Proc. of Annual Conference of
  the International Speech Communication Association (INTERSPEECH)}, 2017, pp.
  1517--1521.

\bibitem{jung2019short}
J.~Jung, H.~Heo, H.~Shim, and H.~Yu, ``Short utterance compensation in speaker
  verification via cosine-based teacher-student learning of speaker
  embeddings,'' in \emph{Proc. of IEEE Automatic Speech Recognition and
  Understanding Workshop (ASRU)}, 2019, pp. 335--341.

\bibitem{jung2020improving}
Y.~Jung, S.~M. Kye, Y.~Choi, M.~Jung, and H.~Kim, ``Improving multi-scale
  aggregation using feature pyramid module for robust speaker verification of
  variable-duration utterances,'' \emph{arXiv preprint arXiv:2004.03194}, 2020.

\bibitem{hochreiter1997long}
S.~Hochreiter and J.~Schmidhuber, ``Long short-term memory,'' \emph{Neural
  computation}, vol.~9, no.~8, pp. 1735--1780, 1997.

\bibitem{mclaren2015softsad}
M.~McLaren, M.~Graciarena, and Y.~Lei, ``Softsad: Integrated frame-based speech
  confidence for speaker recognition,'' in \emph{Proc. of IEEE International
  Conference on Acoustics, Speech and Signal Processing (ICASSP)}, 2015, pp.
  4694--4698.

\bibitem{zhou2019cnn}
T.~Zhou, Y.~Zhao, J.~Li, Y.~Gong, and J.~Wu, ``Cnn with phonetic attention for
  text-independent speaker verification,'' in \emph{Proc. of IEEE Automatic
  Speech Recognition and Understanding Workshop (ASRU)}, 2019, pp. 718--725.

\bibitem{he2016deep}
K.~He, X.~Zhang, S.~Ren, and J.~Sun, ``Deep residual learning for image
  recognition,'' in \emph{Proc. of IEEE Conference on Computer Vision and
  Pattern Recognition (CVPR)}, 2016, pp. 770--778.

\bibitem{vaswani2017attention}
A.~Vaswani, N.~Shazeer, N.~Parmar, J.~Uszkoreit, L.~Jones, A.~N. Gomez,
  {\L}.~Kaiser, and I.~Polosukhin, ``Attention is all you need,'' in
  \emph{Advances in Neural Information Processing Systems}, 2017, pp.
  5998--6008.

\bibitem{ranjan2017l2}
R.~Ranjan, C.~D. Castillo, and R.~Chellappa, ``L2-constrained softmax loss for
  discriminative face verification,'' \emph{arXiv preprint arXiv:1703.09507},
  2017.

\bibitem{warden2018speech}
P.~Warden, ``Speech commands: A dataset for limited-vocabulary speech
  recognition,'' \emph{arXiv preprint arXiv:1804.03209}, 2018.

\bibitem{snyder2015musan}
D.~Snyder, G.~Chen, and D.~Povey, ``Musan: A music, speech, and noise corpus,''
  \emph{arXiv preprint arXiv:1510.08484}, 2015.

\bibitem{paszke2019pytorch}
A.~Paszke, S.~Gross, F.~Massa, A.~Lerer, J.~Bradbury, G.~Chanan, T.~Killeen,
  Z.~Lin, N.~Gimelshein, L.~Antiga \emph{et~al.}, ``Pytorch: An imperative
  style, high-performance deep learning library,'' in \emph{Advances in Neural
  Information Processing Systems}, 2019, pp. 8024--8035.

\bibitem{kingma2015adam}
D.~P. Kingma and J.~Ba, ``Adam: A method for stochastic optimization,''
  \emph{Proc. of International Conference on Learning Representations (ICLR)},
  2015.

\bibitem{lin2017structured}
Z.~Lin, M.~Feng, C.~N.~d. Santos, M.~Yu, B.~Xiang, B.~Zhou, and Y.~Bengio, ``A
  structured self-attentive sentence embedding,'' \emph{Proc. of International
  Conference on Learning Representations (ICLR)}, 2017.

\end{thebibliography}

\end{document}